\documentclass{aa}  
\usepackage[latin1]{inputenc}
\usepackage{graphicx}
\usepackage{longtable}
\usepackage{color}
\usepackage{url}
\usepackage[breaklinks=true,pagebackref=false]{hyperref}
\usepackage{multirow}
\usepackage[table]{xcolor}
\usepackage{supertabular}
\usepackage{txfonts}
\usepackage{natbib}
\usepackage{colortbl}
\bibpunct{(}{)}{;}{a}{}{,} 

\definecolor{gris}{gray}{0.9}

\begin{document}
\title{Phoebe's orbit from ground-based and space-based observations 
}

   \author{J. Desmars
          \inst{1,2}
          \and
          S.N. Li\inst{1,2,3}
          \and
          R. Tajeddine\inst{2,4}
          \and
          Q.Y. Peng\inst{5}
          \and
          Z.H. Tang\inst{1}
          }

 \offprints{J.Desmars, desmars@imcce.fr}

   \institute{Shanghai Astronomical Observatory, Nandan Road 80, The Chinese Academy of Science, Shanghai, 200030, China
         \and
             IMCCE - Observatoire de Paris, UPMC, UMR 8028 CNRS, 77 avenue Denfert-Rochereau,
		75014 Paris, France    
         \and
College of Science, Wuhan University of Science and Technology, Wuhan 430065, China 
	\and	
            Laboratoire AIM, UMR 7158, Universit\'{e} Paris Diderot - CEA IRFU - CNRS, Centre de l'Orme les Merisiers, 91191 Gif sur Yvette Cedex, France 
         \and
             Department of Computer Science, Jinan University, Guangzhou 510632, China
		\\
              \email{desmars@imcce.fr}
             }


 
  \abstract 
   {The ephemeris of Phoebe, the ninth satellite of Saturn, is not very accurate. Previous dynamical models were usually too simplified, the astrometry is heterogeneous and, the Saturn's ephemeris itself is an additionnal source of error. }
   {The aim is to improve Phoebe's ephemeris by using a large set of observations, correcting some systematic errors and updating the dynamical model.}
   {The dynamical model makes use of the most recent ephemeris of planets and Saturnian satellites. The astrometry of Phoebe is improved by using a compilation of ground-based and space-based observations and by correcting the bias in stellar catalogues used for the reduction.}
   {We present an accurate ephemeris of Phoebe with residuals of 0.45 arcsec and with an estimated accuracy of Phoebe's position of less that 100 km on 1990-2020 period.}
   {}

   \maketitle
%

\section{Introduction}

Phoebe is the biggest irregular satellite of Saturn (220km in diameter). It has a retrograde orbit with an inclination of about 176 degrees on the eclipitic plane orbiting at about 13 million km (0.086 au) from Saturn. The ephemeris of Phoebe is not very accurate for three reasons. The main reason is the astrometry of Phoebe. Because of its faintness (about magnitude 16), Phoebe is not easy to observe and was especially difficult to observe in the past. Most of the available observations have been realized in the last twenty years. There are only a few observations in the past and they are not very accurate. Another reason is that the dynamical models of Phoebe's motion are usually not complete and take into account only a few perturbations. However, while the astrometry is not good, a simple dynamical model could be enough to compute the position of Phoebe. The last reason is the dependance on Saturn's position. When comparing the observed and computed positions of Phoebe, the position of Saturn, or more exactly the barycentre of the Saturnian system is required because the dynamical model provides the position of Phoebe in relation to the Saturnian barycentre. Consequently, the ephemeris of Saturn can be the source of systematic error in Phoebe's position.

In this paper, we develop a new ephemeris of Phoebe by improving the three previous limiting factors. In particular, we first develop a new flexible dynamical model that can take into account the perturbations of the planets, the main Saturnian satellites, and the flatness of Saturn, using the most recent planetary and satellite ephemerides (Sect.~\ref{S:Model}). 

An important step is realized for astrometry. We compile a large part of ground-based observations and include also observations from Voyager and Cassini spacecrafts (Sect.~\ref{S:obs}). Moreover, we study old observations and try to reduce some old observations with modern techniques (Sect.~\ref{Ss:reducpickering}). In addition, most of the observations (78\%) are also corrected of bias in the stellar catalogue used for their reduction (Sect.~\ref{Ss:bias}). 

After presenting the fitting process (Sect.~\ref{S:fitting}), we compare four different models of Phoebe's motion (Sect.~\ref{S:comparison}) and highlight the one with the best residuals (Sect.~\ref{S:ephemeris}). Finally, by using statistical methods, we also compute the estimated accuracy of Phoebe's position during the 1875-2020 period (Sect.~\ref{S:accuracy}).


\section{A flexible dynamical model}\label{S:Model}
According to \cite{Jacobson1998}, the previous dynamical models usually took into account the perturbation of some planets. \cite{Rose1979} includes only the Sun's perturbation, \cite{Bykova1982} added Jupiter's perturbation and \cite{Bec1982} also added the perturbation of Titan. For more recent models, \cite{Jacobson1998} took into account the perturbations of the Sun, Jupiter, and Uranus, as well as the perturbations of the main Saturnian satellites for post-1966 observations and the perturbation of an orbiting Titan for pre-1966 observations. \cite{Shen2005,Shen2011} included the perturbations of the Sun, Jupiter and Uranus and the flatness ($J_2$) of Saturn, whereas \cite{Emelyanov2007} took into account the perturbations of the Sun, Jupiter, Uranus, Neptune and Saturn's flatness ($J_2,J_4$). For the two last models, the Saturnian satellites are taken into account by adding their masses to the mass of Saturn and by correcting the $J_2$ and $J_4$ values for \cite{Emelyanov2007}. Finally, \citet{Jacobson2006} proposed a complete model of the Saturnian system by determining the motion of all Saturnian satellites including Phoebe.

\begin{figure}[h!]  
\centering 
\includegraphics[width=\columnwidth]{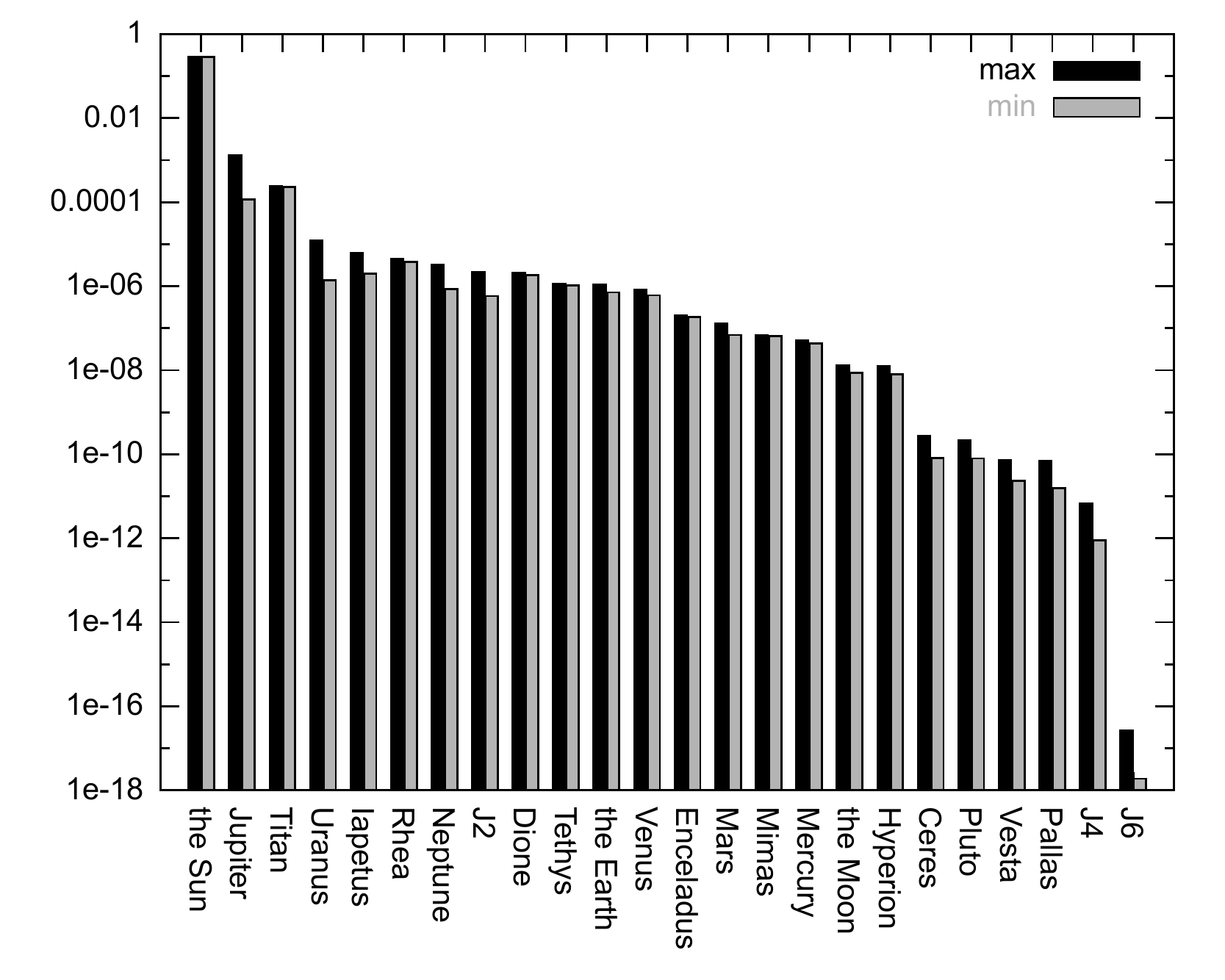}
\caption{Comparison of the maximum and minimum magnitude of gravitational perturbations compared to Saturn's acceleration}\label{F:compforce}
\end{figure}

Figure~\ref{F:compforce} represents the ratio of maximum and minimum magnitude of gravitational accelerations compared to Saturn's acceleration. The main perturbers of Phoebe's motion are the Sun, Jupiter, Titan, Uranus, and Iapetus. Taking into account the flatness of Saturn does not seem justified since other perturbations like those due to the main Saturnian satellites are not included. In particular, the perturbation due to $J_4$ is less important than the gravitational perturbations of the main asteroids.

To measure the effect of perturbers on Phoebe's motion, we have developed a \emph{flexible} model. This dynamical model is adjustable, in the sense that it is possible to take into account or not the perturbations of the Sun, the eight planets and the Moon, the eight major Saturnian satellites and the flatness of Saturn ($J_2$). The masses of the planets are from \cite{DE421} and the other parameters of the Saturnian system are from \cite{Jacobson2006} and are provided in Table~\ref{T:saturnsystem}. 

If a planet is taken into account, it is considered as a point mass and its position is from JPL Planetary Ephemeris DE421 \citep{DE421}. In the opposite case, the mass of the planet is added to the Sun's mass. Likewise, if a Saturnian satellite is taken into account, it is considered as a point mass and its position is given by  \cite{Lainey2012} and in the opposite case, its mass is added to Saturn's mass.

For our model, we use a numerical integration of the equations of motion using a Gauss-Radau integrator \citep{Everhart1985}. These equations are written in the Saturnian system barycentre. In order to fit the dynamical model, the equations of variation are also integrated with the equations of motion \citep[as in][]{Lainey2004}. 

\begin{table}[h!]
\begin{center}
\caption{Dynamical constants of Saturn's system \citep{Jacobson2006}}
\label{T:saturnsystem}
\begin{tabular}{lrr}
\hline
\hline
\textbf{Name}  & \textbf{Value} & \textbf{Units}\\
\hline
 Phoebe GM 				&    0.5534		&  km$^3$.s$^{-2}$ \\
 Mimas GM 				&    2.5023		&  km$^3$.s$^{-2}$ \\
 Enceladus GM				&    7.2096		&  km$^3$.s$^{-2}$ \\  
 Tethys GM				&    41.2097 		&  km$^3$.s$^{-2}$ \\
 Dione  GM				&    73.1127 		&  km$^3$.s$^{-2}$ \\
 Rhea	GM				&    153.9416 		&  km$^3$.s$^{-2}$ \\
 Titan GM 				&    8978.1356 		&  km$^3$.s$^{-2}$ \\
 Hyperion GM				&    0.3727 		&  km$^3$.s$^{-2}$ \\
 Iapetus GM				&    120.5117 		&  km$^3$.s$^{-2}$ \\
 Saturn equatorial radius 		&   60330  		&  km\\
 Saturn $J_2$ $(\times 10^6)$ 		&  16290.71   		&   \\
 Saturn pole $\alpha_P$  		&  40.5955   		&  degrees \\
 Saturn pole $\delta_P$ 		&  83.5381   		&  degrees \\
 Saturn polar rate $\dot{\alpha}_P$ 	&  -0.04229   		&  deg.century$^{-1}$ \\
 Saturn polar rate $\dot{\delta}_P$ 	&  -0.00444  		&  deg.century$^{-1}$ \\
\hline
\end{tabular}
\end{center}
\end{table}

As a large part of astrometric observations is given in absolute coordinates and not in relative coordinates compared to another satellite as for the other main Saturnian satellites, the computed position of Phoebe depends on the computed position of Saturn. Consequently, the position of Saturn can become a source of systematic error in the determination of Phoebe's orbit. 

In particular, some differences in the position of Saturn between several planetary ephemeris exist. In previous models of Phoebe, JPL ephemeris DE406 was usually used. Compared to the most recent ephemeris, DE421, the difference can reach 0.15 arcsec during the 1900-2020 period. In comparison, the difference between INPOP10a \citep{Fienga2011} and DE421 is less than 2 mas during the same period. In these recent ephemerides, the orbit of Saturn is determined using Cassini tracking and ground-based astrometry. For DE421, the accuracy of Saturn's position is tens of km \citep{DE421}. 


\section{Observations of Phoebe}\label{S:obs}
\subsection{Ground-based observations}\label{Ss:Gobs}

The ground-based observations of Phoebe come from four different sources:

\begin{itemize}
\item \textbf{NSDC}: the Natural Satellite Data Center\footnote{NSDC observations of Phoebe are available on this website : \url{http://www.imcce.fr/hosted_sites/saimirror/bsapooue.htm}.} provides a large number of observations of natural satellites \citep{Arlot2009}. For Phoebe, 3404 observations are available from 1904 to 2011 concerning almost 120 different references;
\item \textbf{MPC}: the Minor Planet Center provides observations of natural outer irregular satellites of the giant planets\footnote{\url{http://www.minorplanetcenter.net/iau/ECS/MPCAT-OBS/MPCAT-OBS.html}}. As of 19 September 2012, 2040 observations of Phoebe from 1898 to 2012 are available in the database. The old observations before 1981, usually provided in another frame (such as B1950.0) have been rotated to the J2000. All the observations are given in the J2000 reference frame. Most of the observations come from CCD (about 90\%) and 5 observations made on 13 and 14 June 2010 come from the WISE spacecraft (see Sect.~\ref{S:Sobs});
\item \textbf{NOFS}: 205 NOFS observations \citep{NOFS} from the Flagstaff Observatory were obtained from 2000 to 2011. All the data are available on the web site of the FASTT Planetary Satellite Observations \texttt{http://www.nofs.navy.mil/data/plansat.html};
\item \textbf{Pickering}: \cite{Pickering1908} published 42 observations of Phoebe made in Arequipa, the Harvard observatory station from 1898 to 1904. The observations are provided in B1875 reference frame. They represent the oldest observations of Phoebe. The time of observation is given without indication of seconds, i.e. to the nearest minute. A specific study of these observations has been done in order to improve their astrometry (Sect.~\ref{Ss:reducpickering}). 
\end{itemize}

Many observations appear in several databases. In particular, about 1950 appear in both MPC and NSDC. In that context and as a general rule,  we favour observations in NSDC because they are given in their original format (similar to the publication) whereas MPC turns the observations into J2000 reference system with their own routine. Consequently, only 83 observations from MPC (and only in that database) have been selected. 

In the same way, some observations appear in both the NSDC and the NOFS databases. In particular, observations from \cite{Stone2000,Stone2001} and observations from NOFS compiled in NDSC (so0023 and so0029) are replaced by NOFS observations \citep{NOFS} that use a more recent reduction process and are provided with more digits for time and coordinates. This is also the case of MPC observations provided by the NSDC database (MPC63362, MPC63363, MPC67130, MPC75144, MPC75145) replaced in our compilation by \cite{NOFS}.

Finally, there are also several doubles in the NSDC database. Some observations appear in several references but with different digits for time, or coordinates. In this context, we keep the observations with more digits. For example, observations in MPC52887 are replaced by those of NSDC (so0022 file). Observations in \cite{Mulholland1980} appear twice in the NDSC database (so0001 and so0019) and we keep those in \cite{Strugnell1990} (so0001).

\subsection{Space-based observations}\label{S:Sobs}
During its encounter with Saturn in 1981, Voyager 2 performed eight imaging observations of Phoebe. These observations from 17 June to 15 August 1981 were published in \cite{Jacobson1998}. They are provided in specific coordinates which are pixel and line locations in the Voyager camera frame. The Saturn barycentric position and the velocity of Voyager 2 during the eight observation times are also provided in the paper. The process of reduction that takes into account the camera pointing, the gnomonic projection, the electromagnetic and optical distortion, is fully described in \cite{Jacobson1998}. In this paper, we also use pixel and line locations in the camera frame as coordinates and the positions of Voyager given in \cite{Jacobson1998} are also used. \\

The Cassini spacecraft entered Saturn's orbit at the end of June 2004. Ever since, its ISS Narrow Angle (NAC) and Wide Angle (WAC) cameras have sent tens of thousands of images of Saturn, its rings, and its satellites. Right before entering Saturn's orbit, Cassini performed a Phoebe flyby on 11 June 2004, where it reached a minimum distance of about 2068 kilometers. Cassini's ISS NAC kept observing that satellite from 6 to 12 June 2004, producing hundreds of high-resolution images of Phoebe that can be downloaded from Planetary Data System\footnote{\url{http://pds.nasa.gov/}}. 

A total number of 223 images from Phoebe's flyby were used for astrometric reduction. The spacecraft's position and camera's pointing vector were computed using SPICE library\footnote{\url{http://naif.jpl.nasa.gov/naif/}} \citep{Acton1996}. Since there is an error in the camera's pointing vector, its correction was done using the UCAC2 star catalogue \citep{Zacharias2004}. The satellite's position measurement was done using the ellipse projection of Phoebe's triaxial shape \citep{Thomas2010}. Nonetheless, a difficulty was encountered for the centre-of-figure measurement since Phoebe has an irregular shape and fitting an ellipse is not the best way to determine its position. The astrometric model that was used for pointing correction and satellite centre-of-figure measurement is well described in \cite{Tajeddine2013}. The number of detected stars varied from 2 to 43 stars per image, depending on the satellite's size on the image and camera exposure length. 
	
\begin{figure}[h]
\centering\includegraphics[width=\columnwidth]{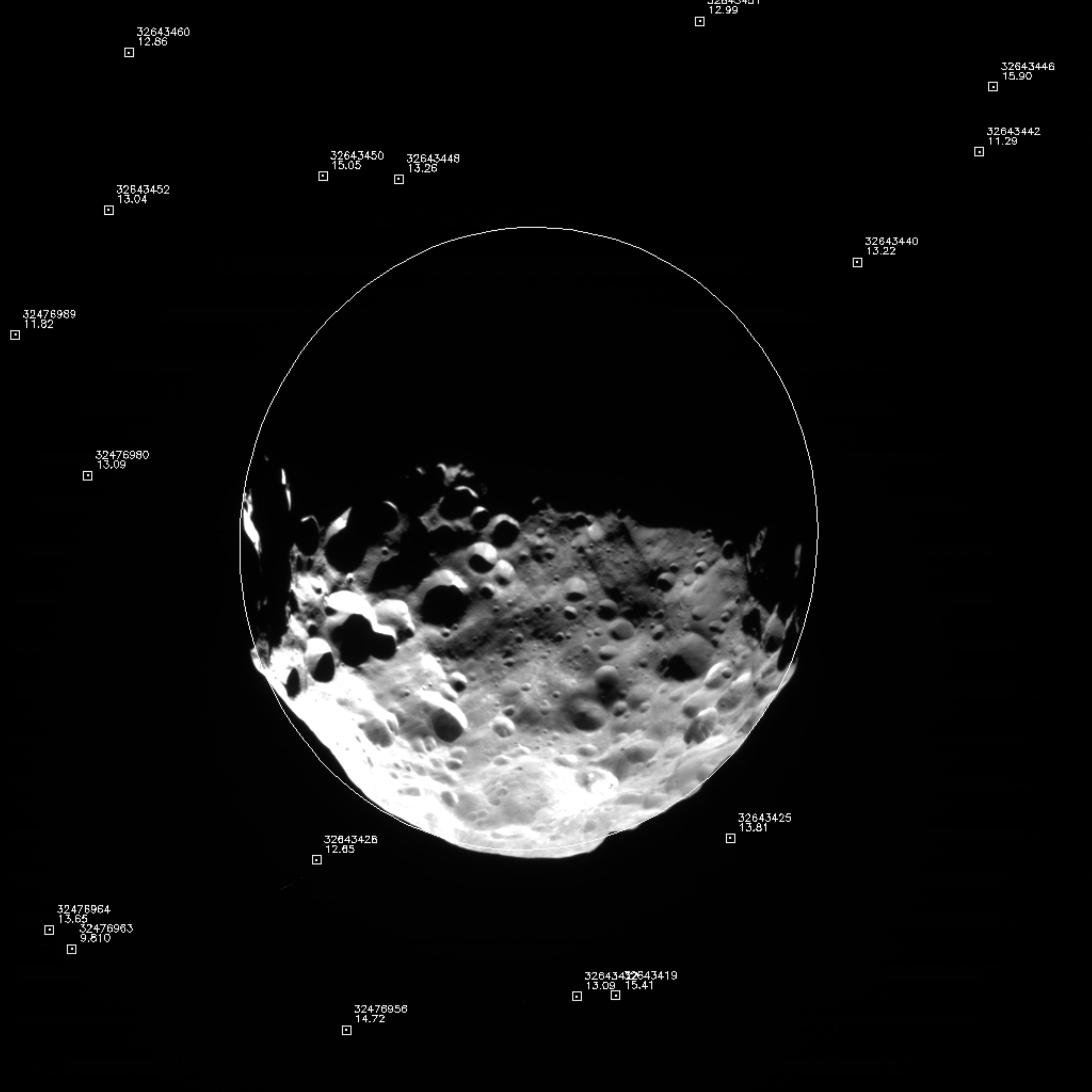}
\caption{Example of a Cassini astrometric reduction of Phoebe. UCAC2 catalogue stars are superimposed on the image.}\label{F:cassini}  
\end{figure}

Figure~\ref{F:cassini} shows an example of Phoebe's astrometric reduction from the Cassini spacecraft and the resolution reached. With an obtained accuracy of a few kilometers, the Cassini observations have proven to be critical in Phoebe orbit modelling.\\

Finally, the WISE spacecraft (Wide-Field Infrared Survey Explorer) also provided five more space-based observations in 2009. But unlike to Cassini or Voyager 2, this spacecraft was in Earth-orbit and not in the vicinity of Saturn system. These observations can be treated as ground-based observations.

\subsection{Catalogue of observations of Phoebe}
A total of 3598 observations (3367 ground-based and 231 space-based) from 1898 to 2012 have been compiled in a catalogue. For comparison, in previous studies \cite{Emelyanov2007} used 1606 observations from 1904 to 2007 and \cite{Shen2011} used 2994 observations from 1904 to 2009.

For the catalogue of observations, we adopted the format presented in \cite{Desmars2009a} and we added some information about the catalogue used for astrometric reduction (using the MPC flag), the correction applied on observations, and the weight used for the fit. For specific observations, we also add the year of epoch or the angular parameters of the Voyager camera. The full catalogue can be requested from the authors. 

\begin{figure}[h!]  
\centering 
\includegraphics[width=\columnwidth]{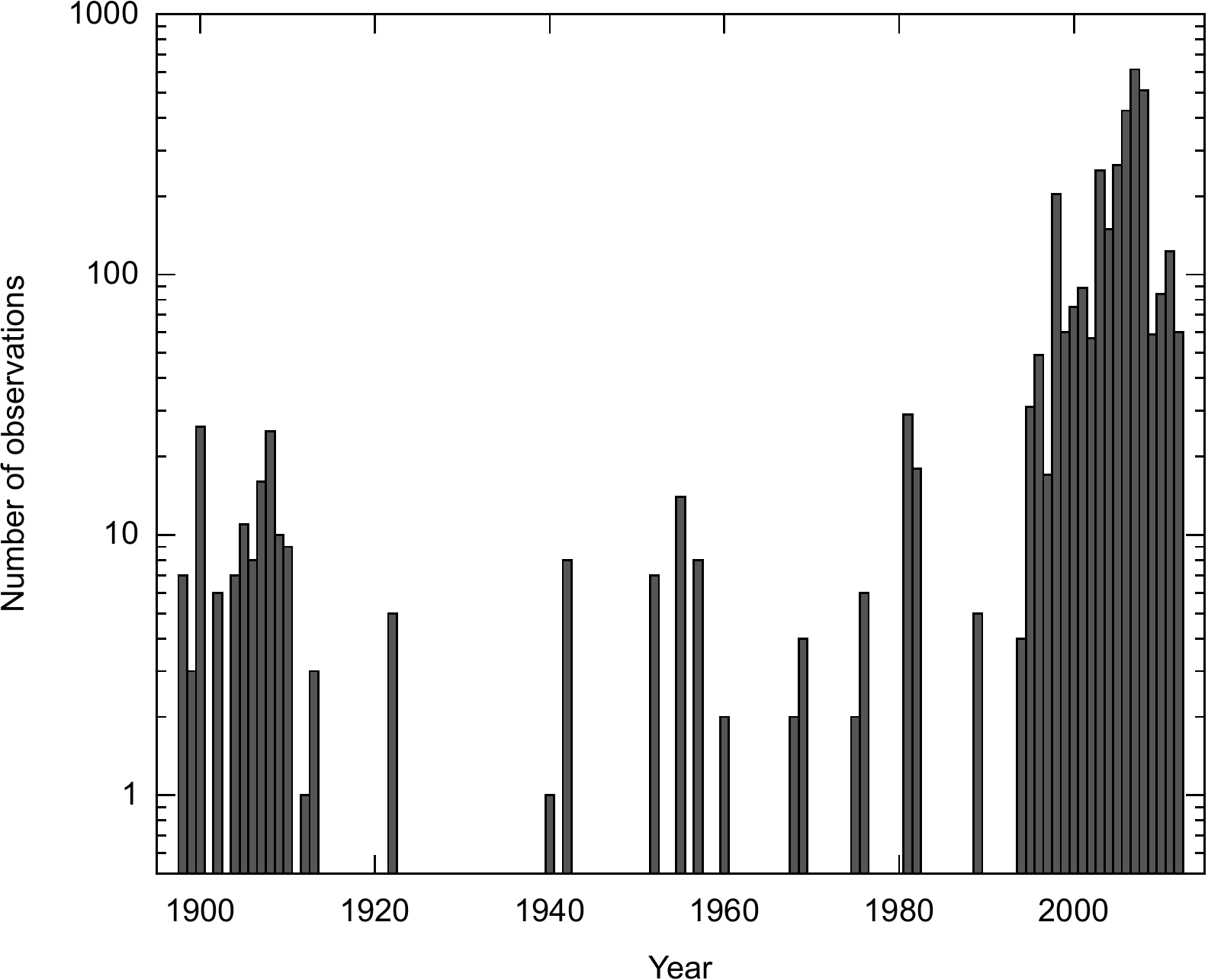}
\caption{Logarithmic distribution of the Phoebe observations}\label{F:repobs}
\end{figure}

Figure~\ref{F:repobs} represents the distribution of Phoebe observations and reveals its strong sparsity. Most of the observations have been realized since the mid-1990s. Other observations were from the early 20th century and there are a few data between these two periods. This strong disparity is a problem for determining the accurate orbit of Phoebe.  

\section{Improvement of astrometry}\label{S:improvement}
In this section, we try to improve the astrometry of Phoebe in two ways. The first is a new reduction of old data, and the second is the correction of bias in stellar catalogues.

\subsection{New reduction of old observations}\label{Ss:reducpickering}
\begin{figure}[h!]  
\centering 
\includegraphics[width=\columnwidth]{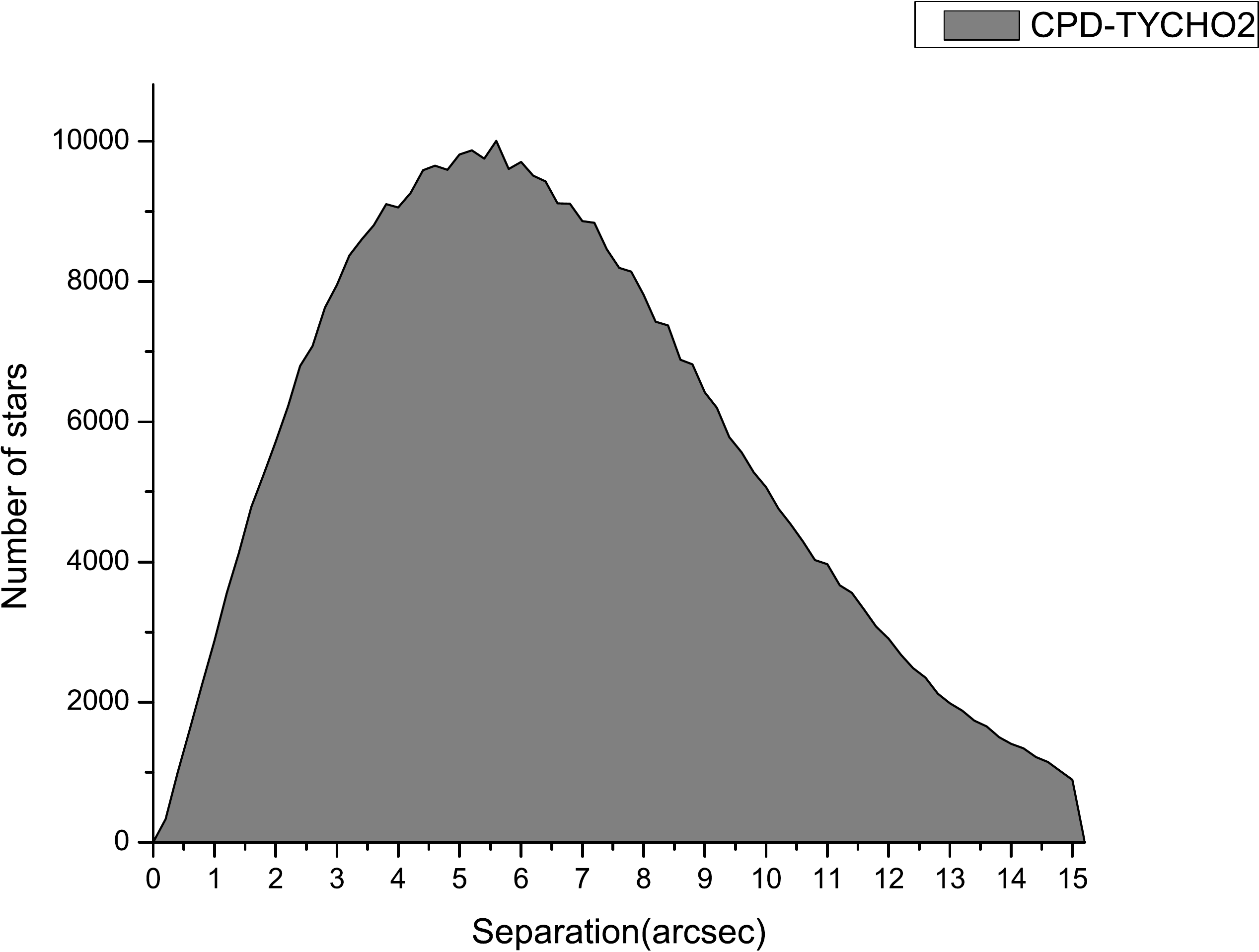}
\includegraphics[width=\columnwidth]{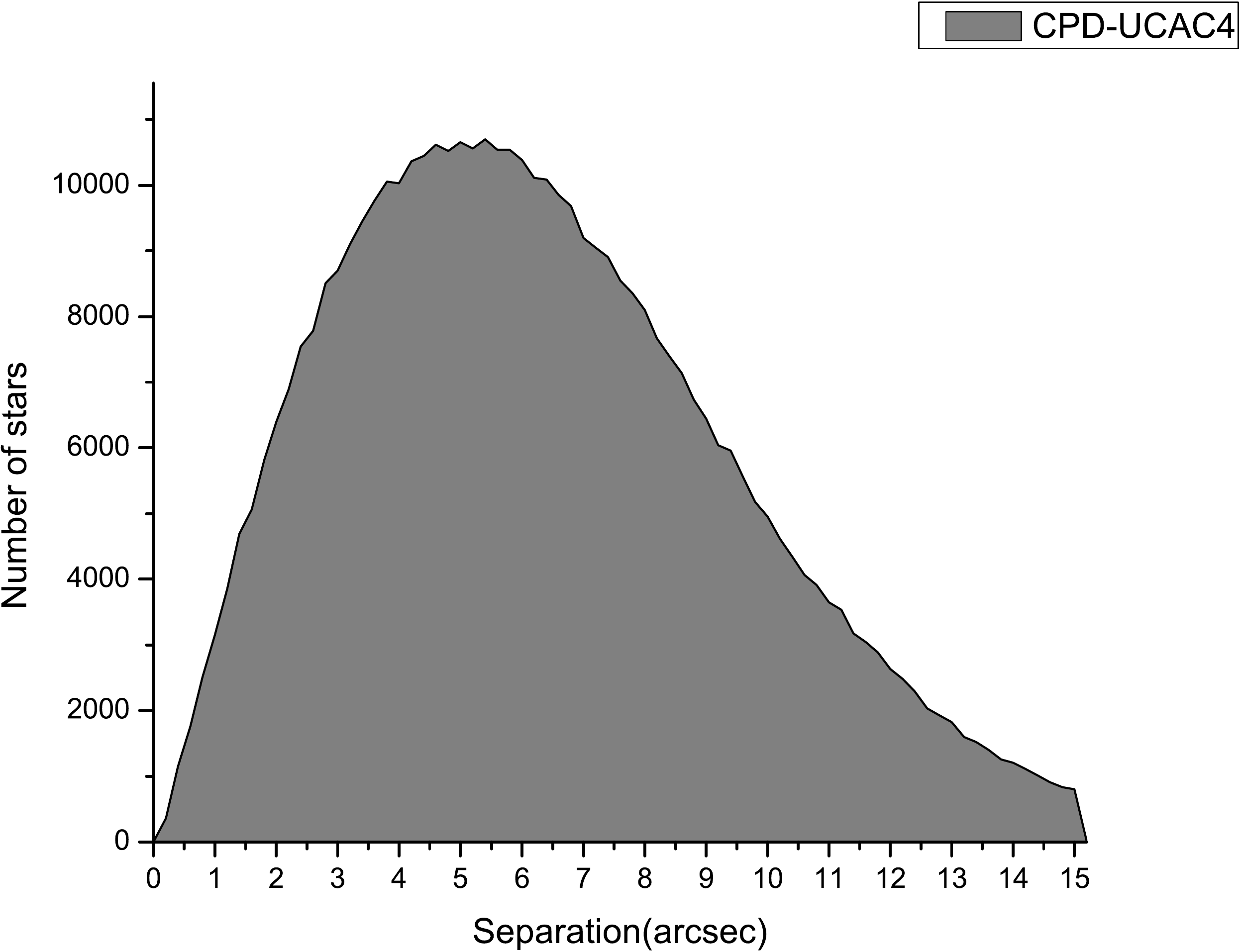}
\caption{Distribution of the separation of pairs of stars in the Tycho2-CPD catalogues (top) and UCAC4-CPD catalogues (bottom).}\label{F:catalogues}
\end{figure}

In some older publications such as \cite{Pickering1908} and \cite{Perrine1904}, authors gave the spherical positions of the natural satellite and its relative positions to reference stars on the photographic plates. In their papers, the position of Phoebe is deduced from the positions of reference stars of the Cape Photographic Durchmusterung (CPD) catalogue. Unfortunatly, the number of reference stars of this catalogue on photographic plates is small and their positions are inaccurate. As an illustration, Fig.~\ref{F:catalogues} represents the statistics of the difference in angular separation\footnote{In this context, angular separation is given by $s=\sqrt{\Delta \alpha^2 \cos^2 \delta + \Delta \delta ^2}$ where $\Delta \alpha$ and $\Delta \delta$ are the difference between right ascension and declination given by older and modern catalogues.} of the stars in CPD and modern catalogues used in current reduction. For many stars, the difference in position between the older and recent catalogues is more than 5 arcsec. The positions of the reference stars in the CPD catalogue represent a source of systematic errors on Phoebe's positions. 

Nowadays, although these plates are not available, it is possible to reduce the positions of Phoebe from data provided by these publications with modern and precise astrometric catalogues such as TYCHO and UCAC4 \citep{Zacharias2013}. The authors gave the position of Phoebe relative to the position of the reference stars in CPD catalogue. With accurate positions of reference stars given in the modern catalogues, it is possible to deduce the spherical position of Phoebe on these plates. The method of reduction consists in identifying the reference stars on the plates and applying corrections of proper motion for each star at the date of observation. In the identification process, two stars in two different catalogues are considered the same star if the distance between their position is smaller than 15 arcsec and if the difference of their magnitude is less than one. 

\begin{figure}[h!]  
\centering 
\includegraphics[width=\columnwidth]{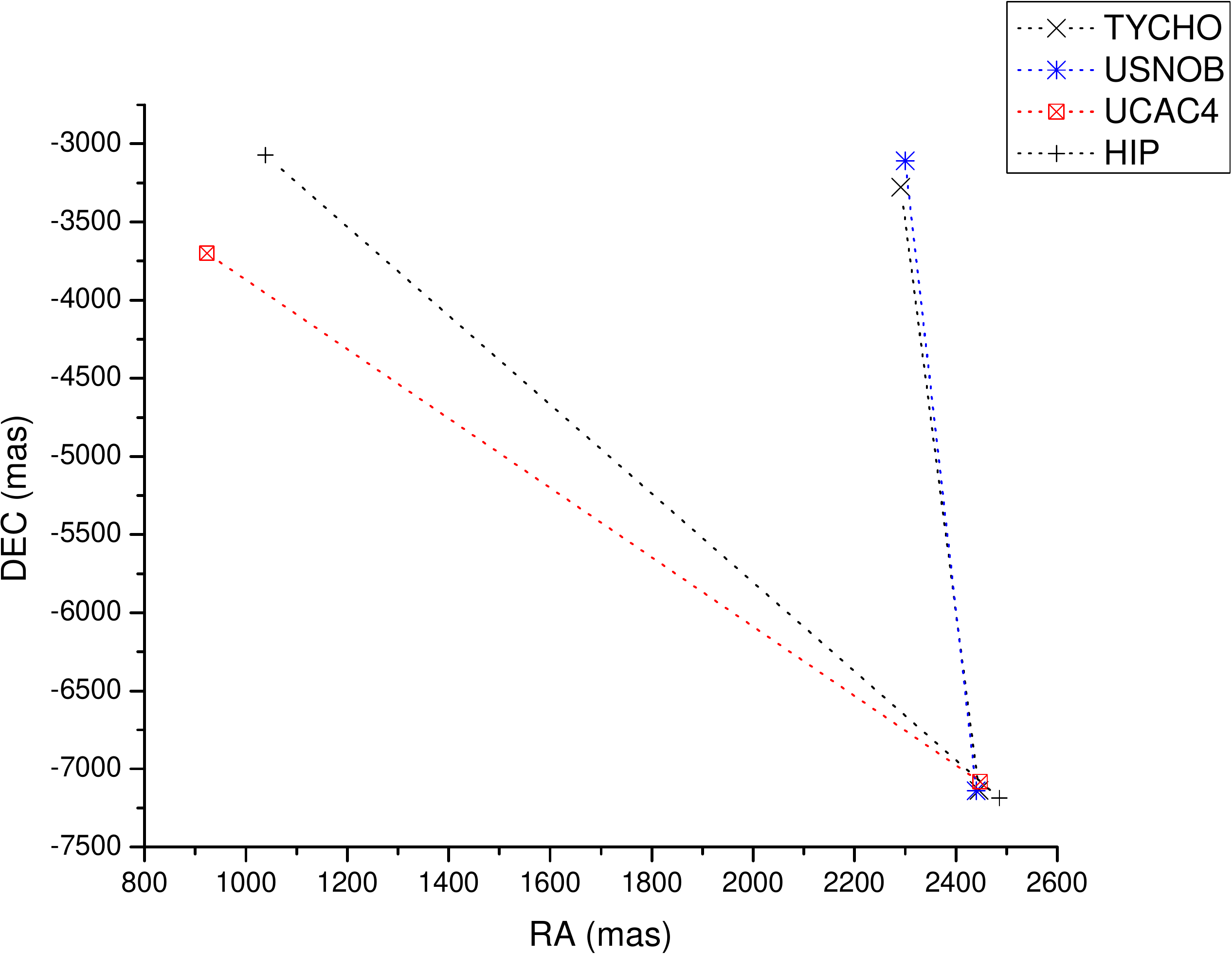}
\caption{The positions of one reference star in epoch J2000.0 (lower) and B1900.0 (upper) calculated with the positions and the proper motions in four different catalogues: TYCHO, USNOB1.0, UCAC4, and HIPPARCOS.}\label{F:proper_motion}
\end{figure}

We tried to reduce observations from \cite{Pickering1908} with this method and by using several stellar catalogues (TYCHO, USNOB, and UCAC4). The residuals obtained are at the same level of accuracy as the positions provided by Pickering (2-3 arcsec). In fact, the main difficulty of this method is the poor knowledge of the proper motion of reference stars. The observations were realized more than 100 years ago and consequently inaccurate values of proper motion lead to inaccurate positions of stars at the date of observation. For example, Fig.~\ref{F:proper_motion} represents the position of a reference star given by four different catalogues at J2000.0 and B1900.0 epochs using the star proper motions from these catalogues. For the J2000 epoch, the positions of the star are quite similar for the different catalogues. For B1900.0, the difference in the positions of the star can reach more than 1 arcsec, according to the catalogue used. The HIPPARCOS catalogue is expected to give accurate values of proper motions but only a small number of stars from this catalogue is available on the plate (less than four).

The inaccurate values of proper motion of stars are currently a difficulty in the new reduction of old data. Finally, for the Pickering and Perrine observations, we used the positions provided in the papers. A possible solution would be given by the Gaia space mission and its astrometric stellar catalogue. Gaia will provide accurate positions and accurate proper motions of stars \citep{Mignard2007}. In that context, the method described in the section could be successfully applied. 

\subsection{Stellar catalogue correction}\label{Ss:bias}
Stellar catalogues used in astrometric reduction are usually a source of systematic error. This error depends on the catalogue used and on the zone on the celestial sphere.

\cite{Chesley2010} propose a treatment of star catalog biases in asteroid astrometric observations. Considering the 2MASS stellar catalogue as the reference, they have detected a bias in star positions of five stellar catalogues (Tycho-2, UCAC2, USNO B1.0, USNO A2.0, USNO A1.0), deduced from the residuals of numbered asteroid observations. They also provided a method to remove the biases for observations. 

\begin{figure}[h!]  
\centering 
\includegraphics[width=\columnwidth]{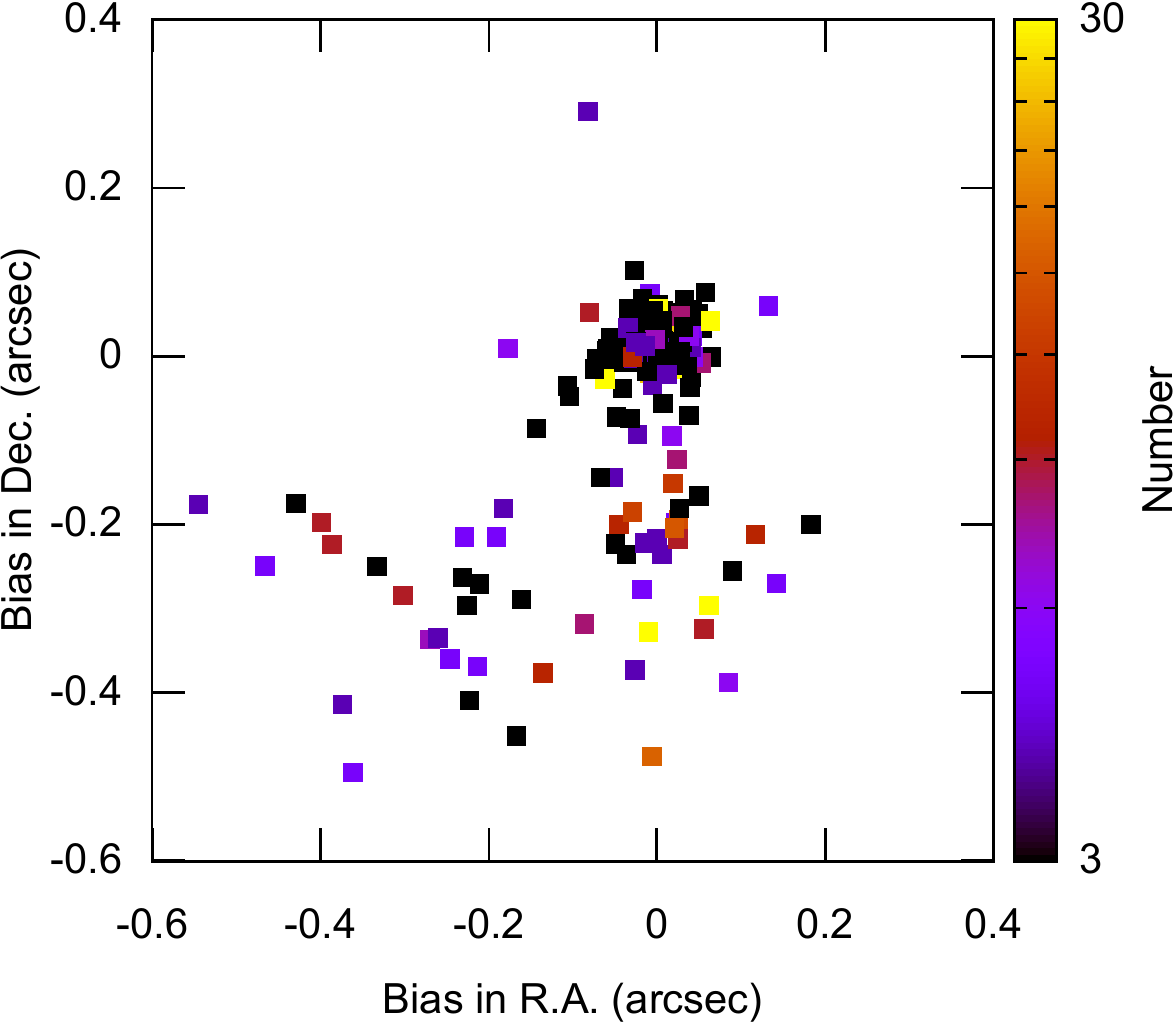}
\caption{Bias in right ascension and declination, and number of observations concerned by bias correction (see text).}\label{F:bias}
\end{figure}

This method can be applied for the first time to a natural satellite such as Phoebe because most of its observations are in absolute coordinates, which is not the case for inner or major satellites (where inter-satellite positions are also used). Table~\ref{T:catalogue} provides statistics on the catalogues used to reduce the astrometric positions of Phoebe. The stellar catalogue can be identified for 2915 observations (86.6\%).  Among these observations, 2625 (78.0\%) use one of the five catalogues studied in \citet{Chesley2010} and are pertained to the treatment. Figure~\ref{F:bias} represents the bias correction applied to Phoebe's observations. As many observations of Phoebe have been realized in the same zone on the celestial sphere, these observations have the same bias correction. Consequently, the number of observations for a specific bias correction is also indicated in Fig.~\ref{F:bias}. The bias removal can reach about 0.5 arcsec both in right ascension and in declination.

\begin{table}[h!]
\begin{center}
\caption{Statistics on catalogues used for the reduction of the Phoebe observations. Code is similar to MPC flag.}
\label{T:catalogue}
\begin{tabular}{clrrc}
\hline
\hline
\textbf{Code}  & \textbf{Catalogue}  & \textbf{Number} & \textbf{P0ercentage} & \textbf{Time-span} \\
\hline
a & USNO A1.0            &     8 &   0.2\% & 2000-2000\\
b & USNO SA1.0           &     3 &   0.1\% & 2000-2000\\
c & USNO A2.0            &   384 &  11.4\% & 1998-2012\\
d & USNO SA2.0           &    12 &   0.4\% & 2001-2003\\
g & Tycho-2              &   236 &   7.0\% & 2000-2011\\
l & ACT                  &     5 &   0.1\% & 2000-2000\\
o & USNO B1.0            &   272 &   8.1\% & 2005-2012\\
r & UCAC2                &  1725 &  51.2\% & 1996-2012\\
t & UCAC3-beta           &     6 &   0.2\% & 2011-2012\\
u & UCAC3                &    76 &   2.3\% & 2010-2012\\
v & NOMAD                &    95 &   2.8\% & 2008-2009\\
w & CMC                  &     2 &   0.1\% & 2010-2010\\
z & GSC (generic)        &    27 &   0.8\% & 2000-2000\\
  & Unknown              &   516 &  15.3\% & 1898-2010\\
\hline
\end{tabular}
\end{center}
\end{table}

\section{Fitting process}\label{S:fitting}

\subsection{Weighting process}
The fitting process consists in the determination of the epoch state vector of Phoebe (components of position and velocity at a given epoch) that minimizes the residuals (observed position minus computed position). The classic least-squares method (LSM) is used for this determination \citep[for more information, see for example][]{Desmars2009b}. In the LSM, a weighting matrix $V_{obs}$ is required and usually considered as a diagonal matrix where the diagonal components are $\epsilon_i^2=1/\sigma_i^2$ where $\sigma_i^2$, is the estimated variance of the observation $i$. 

For our process, we first fit the model to all observations by considering their weight as identical and $\sigma=1$ arcsec for each observation. Then, the root mean square (rms) of the O-C in both coordinates (right ascension and declination) is computed for each reference and each observatory. Finally, we fit again the model by considering these rms depending on reference and observatory as their weight.  

\subsection{Correction applied to observations}

In the fitting process, observations can be corrected for different bias. In particular, as described in Sect.~\ref{Ss:bias}, correction of stellar catalogue bias can be corrected. Moreover, for prior 1940 observations, a correction in right ascension of 0.75 arcseconds should be applied as explained in \cite{Jacobson2000}.

\subsection{Change of reference frames}
Most of observations of Phoebe are provided in absolute coordinates (right ascension and declination) but in a different reference system (ICRF, J2000, B1950, true equator and equinox of the date, mean equator and equinox of the date, mean equator and equinox at 1st January of the year of observation, mean equator and equinox at 1st January of a specific year, mean equator and equinox of specific epoch).

Because the dynamical model allows the computation of the positions of Phoebe to ICRF, we have to transform observations in ICRF in order to compare computed and observed positions. The classical transformations have been done with the SOFA routine \citep{SOFA}. For transformation from B1950, the technique from \cite{Murray1989} is applied. 

The case of observations from \cite{Pickering1908} published in B1875 is particular. The observations are first corrected from elliptical aberration, then transformed to mean equator and equinox of the date using the precession value of Newcomb \citep{Kinoshita1975}. Finally, the coordinates are transformed to ICRF by using current values of precession (IAU Resolutions 2006).


\section{Comparison of dynamical models}\label{S:comparison}

The flexible model developed in Sect.~\ref{S:Model} allows the definition of four different models of the motion of Phoebe:
\begin{itemize}
\item \textbf{Model 1} is the most complete version of the flexible model with all the perturbations: gravitational perturbations of the Sun, of all the planets and the Moon, then the perturbations of the eight main satellites of Saturn, and the flatness parameter $(J_2)$; 
\item \textbf{Model 2} includes the perturbations of the Sun, Jupiter, Uranus and Neptune, then the three main Saturnian satellites (Titan, Iapetus, Rhea), and the $J_2$ parameter; 
\item \textbf{Model 3} includes only perturbations of the Sun, Jupiter, and Titan;
\item \textbf{Model 4} includes the perturbations of the Sun and Jupiter. 
\end{itemize}

The aim of the four models is to determine if a full model is required to compute the position of Phoebe or if a simple model is enough. In that context, we compute the residuals of the Phoebe observations after fitting to the four models. Table~\ref{T:rms} provides the mean and the root mean square of the O-C for the different models.  If O-C were greater than 5 arcsec, they were rejected. For each case, 3319 observations are accepted among a total number of 3367 ground-based observations and all Cassini data are accepted, representing 223 observations.

\begin{table}[h!]
\begin{center}
 \caption{Mean $\mu$ and root mean square (rms) of O-C in arcsec for the different models.}\label{T:rms}
\begin{tabular}{|cc|r|r||r|r|}
\hline
\hline
\multirow{2}{*}{}
    &	& \multicolumn{2}{c||}{Ground-based obs.} & \multicolumn{2}{c|}{Cassini obs.}\\
\cline{3-6}
    &	& \multicolumn{1}{c|}{$\mu$} & \multicolumn{1}{c||}{rms} & \multicolumn{1}{c|}{$\mu$} & \multicolumn{1}{c|}{rms}  \\
\hline
Model 1 & $\alpha$ &  0.0089  &  0.4467 & 0.6304  &  2.5353 \\
        & $\delta$ &  0.0564  &  0.4617 &-0.1291  &  1.5782 \\
\hline
Model 2 & $\alpha$ &  0.0088  &  0.4468 & 0.6368  &  2.5392 \\
        & $\delta$ &  0.0564  &  0.4617 &-0.1277  &  1.5782 \\
\hline
Model 3 & $\alpha$ &  0.0084  &  0.4514 & 0.6506  &  2.5614 \\
        & $\delta$ &  0.0532  &  0.4612 &-0.1232  &  1.5793 \\
\hline
Model 4 & $\alpha$ &  0.0079  &  0.4675 & 0.6897  &  2.6214 \\
        & $\delta$ &  0.0458  &  0.4649 &-0.1162  &  1.5813 \\
\hline
\hline
\end{tabular}
\end{center}
\end{table}

The residuals for the four models are quite similar in particular for Models 1 and 2. As expected, the perturbations included in the first model but not in the second\footnote{Gravitational perturbations of Dione, Tethys, Earth, Venus, Enceladus, Mars, Mimas, Mercury, the Moon, and Hyperion.} have no main influence on the motion of Phoebe. 

The residuals for ground-based observations represent about 4000 km in distance whereas the residuals for Cassini data represents only 4.7 km and provide an important constraint on the motion of Phoebe (see also Sect.~\ref{S:accuracy}).

With regard to Table~\ref{T:rms}, the Model 1 has the best residuals, but a simple model with only the perturbations of the Sun and Jupiter can be enough to compute the motion of Phoebe. Even if the dynamical model is enhanced, there is no clear improvement in the residuals. Although the astrometry is better (large number of observations, correction of bias, addition of space-based observations), it remains unreliable. Consequently, regular observations of Phoebe are still necessary. The method of reduction of old observations presented in Sect.~\ref{Ss:reducpickering} and the future Gaia stellar catalogue also appear to be a good opportunity for the improvement of the astrometry. Even if the difference in the residuals of the four models are not important, the full Model 1 provides the best residuals and we deal with this model hereafter.\\

Table~\ref{T:rmsbias} presents the residuals computed with Model 1 with and without correction of the bias in the stellar catalogue (see Sect.~\ref{Ss:bias}). The correction of the bias can help to reduce the residuals, in particular in declination.

\begin{table}[h!]
\begin{center}
 \caption{Post-fit residuals (mean $\mu$ and root mean square) after correction or not of bias in stellar catalogue. Residuals greater than 5 arcsec were rejected.}\label{T:rmsbias}
\begin{tabular}{|c|c|c||c|c|c|}
\hline
\hline
 & \multicolumn{2}{c||}{Uncorrected} & \multicolumn{2}{c|}{Corrected} & \multirow{2}{*}{Number}\\
\cline{2-5}
 & $\mu$ & rms & $\mu$ & rms & \\
\hline
 $\alpha$ & 0.0241  & 0.4590 & 0.0089  &  0.4467 & 3319/3367\\
 $\delta$ & 0.0857  & 0.4856 & 0.0564  &  0.4617 & 3319/3367\\
\hline
\hline
\end{tabular}
\end{center}
\end{table}


\section{Ephemeris of Phoebe}\label{S:ephemeris}
\begin{figure}[h!]  
\centering 
\includegraphics[width=\columnwidth]{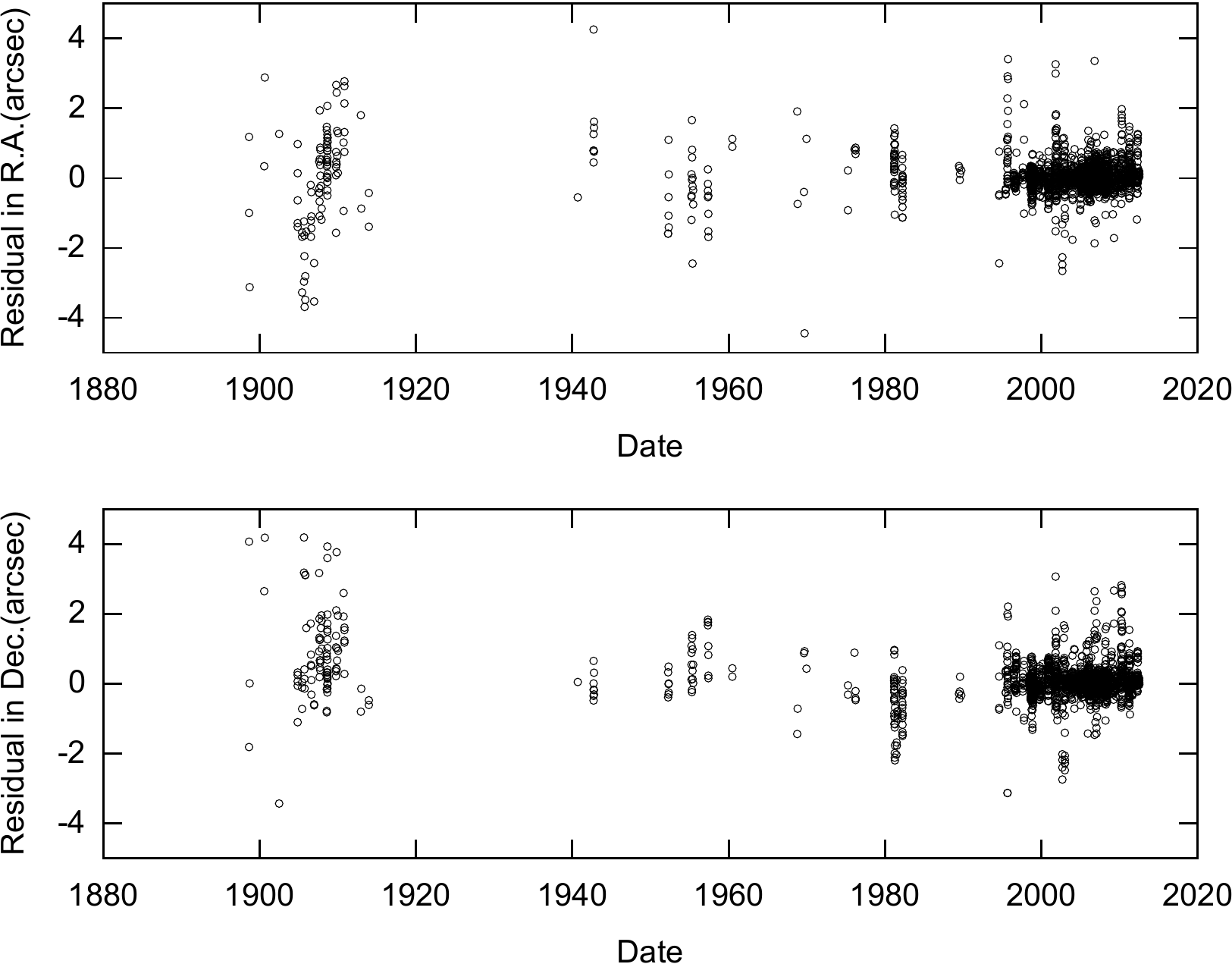}
\caption{Residuals in right ascension (R.A.) and declination (Dec.) of observations}\label{F:omc}
\end{figure}

In Sect.~\ref{S:comparison}, we have shown that Model 1 including the perturbations of planets and satellites, and the flatness parameter provides the best residuals. In this context, Model 1 fitted to the whole set of observations is now considered our Phoebe ephemeris and is called PH12. The statistics of the residuals for each reference is presented in Table \ref{T:statsrms}. Figure~\ref{F:omc} represents the residuals in right ascension and declination for all ground-based observations.

We have derived from the numerical integration a binary file for ephemeris computation using the Chebychev polynomials available at \url{ftp://ftp.imcce.fr/pub/ephem/satel/ph12/}. 
The PH12 ephemeris is available from 1876 to 2022.

\section{Accuracy of the extrapolated position of Phoebe}\label{S:accuracy}
The accuracy of the position of Phoebe can be estimated with statistical methods. \cite{Desmars2009b} presented and developed some statistical methods for estimating the accuracy of satellite ephemerides. \cite{Emelyanov2010} also estimated the ephemeris precision by three different methods for all outer satellites including Phoebe. These methods are Monte Carlo processes and consist in modifying the nominal orbit with specific assumptions. The modified orbit is then compared to the nominal orbit and the process is repeated many times giving a statistical estimation of the accuracy.

In order to estimate the position uncertainty of Phoebe we use two of these methods given in \cite{Desmars2009b}. 
The first, Monte Carlo using Covariance Matrix (MCCM) consists in adding a random noise to observations, whereas the second, bootstrap resampling (BR) consists in randomly sampling the observations. For both methods we generate $K=200$ clones of the nominal orbit.

A measure of the accuracy in distance $\sigma_d$ can be computed with this relation

\begin{equation}
\sigma_d(t)=\sqrt{\frac{1}{K}\sum_{k=1}^K d_k(t)^2},
\end{equation}
where $d_k(t)$ is the distance between the position on the nominal orbit and the position on orbit $k$ at time $t$. \\

\begin{figure}[h!]  
\centering 
\includegraphics[width=\columnwidth]{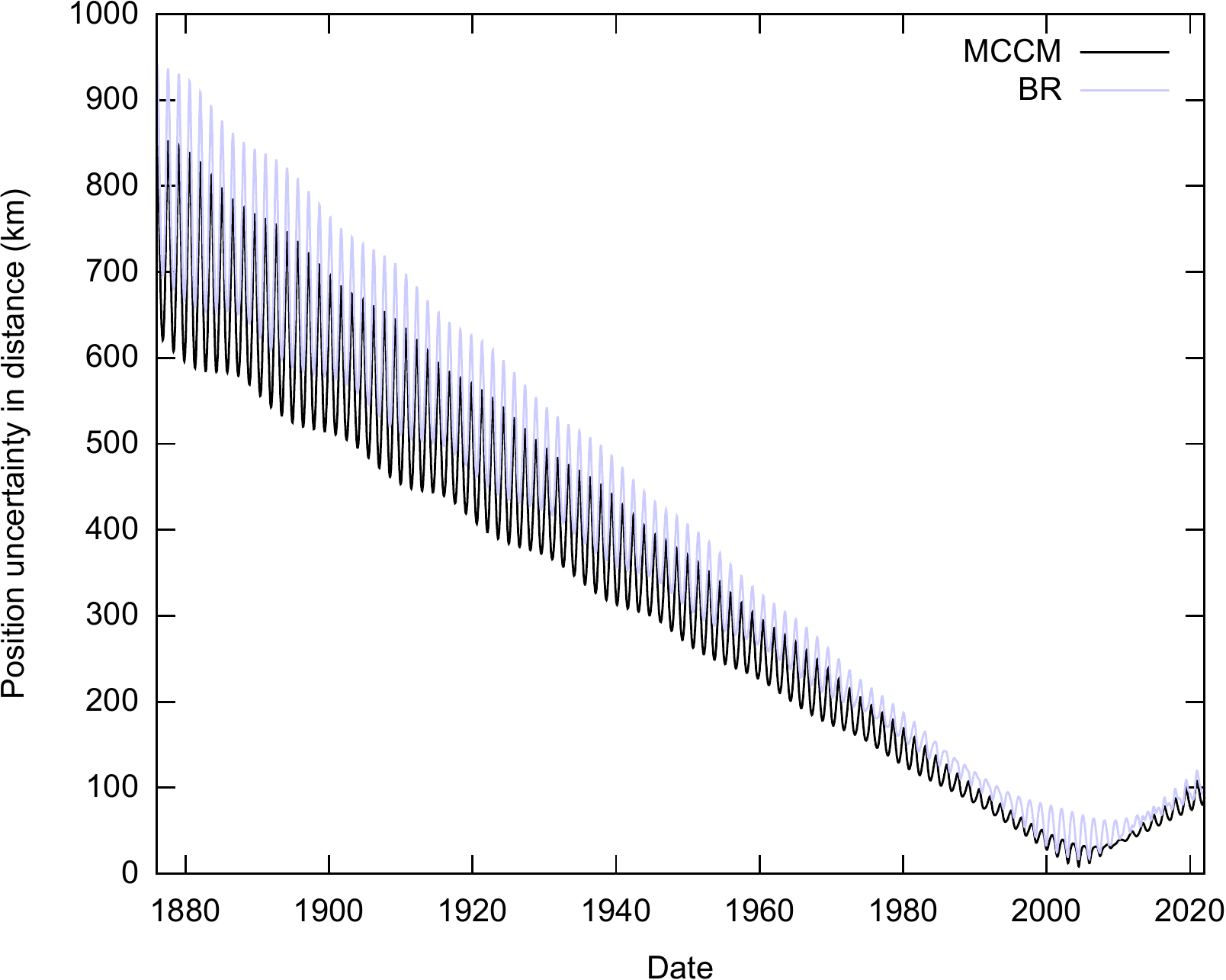}
\caption{Position uncertainty of the Phoebe ephemeris in distance calculated with Monte Carlo using Covariance Matrix (MCCM) and Bootstrap Resampling (BR) as a function of time in years.}\label{F:accuracy}
\end{figure}

The position uncertainty obtained by the two methods is represented in Fig.~\ref{F:accuracy}. Both methods give similar results. The accuracy of the Phoebe ephemeris is quite good during the 2000s thanks to a lot of ground-based observations and, in particular, thanks to Cassini data. During this period, the accuracy is about 50km in distance (which represents about 0.01 arcsec in angular separation) and less than 10km during the Cassini flyby (June 2004). Going back to the past, the accuracy deteriorates owing to the lack of observations and to inaccurate observations. The accuracy reaches about 1000km (about 0.2 arcsec) in the early 20th century. In the near future, the accuracy will remain less than 100km (about 0.02 arcsec). As a comparison, \cite{Emelyanov2010} provides an accuracy of 0.05-0.06 arsec during 2010-2020, but he used only 1606 observations until 2007 and the Cassini data were not fitted.

\section{Conclusion}
In this paper we have developed a flexible dynamical model of Phoebe that includes the perturbations of the Sun, all the planets, the Moon, the main Saturnian satellites, and the flatness parameter. We used the most recent theories of planets and satellites. An important step in the improvement of astrometry has been realized. We extended the catalogue of available observations of Phoebe from 1898 to 2012, then we dealt with the treatment of old observations, and the correction of bias in the stellar catalogue. Finally, we have produced a new ephemeris of Phoebe. The residuals are  0.45 arcsec and the estimated accuracy of Phoebe's position is less than 100km for the 1990-2020 period. The comparison of the different models also shows that a simple model is enough to determine the position of Phoebe, even if the more complete model provides the best residuals. 

The main constraint in the development of an accurate ephemeris of Phoebe is the astrometry. Consequently, this is the point to improve in the future. One way is to regularly observe the satellite. 
Another way is to improve the process of reduction. Recently, \cite{Peng2012} dealt with the geometric distortion in astrometric images and they successfully applied their method to the observations of Phoebe. Finally, the main progress for the Phoebe ephemeris will be achieved with the Gaia space mission. Indeed, Gaia will provide an accurate stellar catalogue (with accurate positions of stars and accurate proper motions). Thus, the method described in Sect.~\ref{Ss:reducpickering} could be effective and provide accurate old data. Moreover, CCD frames and old photographic plates could be reduced again with this stellar catalogue.

\begin{acknowledgements}
The main part of this research work was financially supported by the National Natural Science Foundation of China (Grant N. 11150110138, 10973007, 11273014). The work on the Cassini data was also funded by UPMC-EMERGENCE (contract number EME0911). 
\end{acknowledgements}

\bibliographystyle{aa} 
\bibliography{biblio} 

\clearpage \onecolumn
\begin{small}
\begin{longtable}{rlrrrrrl}
\caption{Statistics of astrometric residuals. Code bib. is the bibliographic code used in the catalogue, $n$ is the number of accepted observations for each reference,  $N$ is the total number of observations of the reference,  $\mu_{\alpha}$ and $\mu_{\delta}$  are the mean of the residuals for each coordinate in arcsec, rms$_{\alpha}$ and rms$_{\delta}$ are the root mean squares for each coordinate in arcsec and, the time-span is also indicated. The symbol * indicates that the observations have been corrected for bias.}\label{T:statsrms}\\
\textbf{Code bib.} & \textbf{Reference} & $n/N$ & $\mu_{\alpha}$ & $\mu_{\delta}$ & rms$_{\alpha}$ & rms$_{\delta}$ & \textbf{Time-span}\\
\hline	
\hline	
\endfirsthead
\caption{continued.}\\
\textbf{code bib.} & \textbf{Reference} & $n/N$ & $\mu_{\alpha}$ & $\mu_{\delta}$ & rms$_{\alpha}$ & rms$_{\delta}$ & \textbf{time-span}\\
\hline	
\hline	
\endhead
\multicolumn{8}{c}{\emph{Observations in \cite{Strugnell1990}}} \\
 13 & \cite{Chernykh1971}                &     2/    2 &     0.5854 &	 -1.0816 &     1.4508 &     1.1405 & 1968-1968  \\
 34 & \cite{Mulholland1980}              &     6/    8 &     0.4136 &	 -0.0970 &     0.7573 &     0.4710 & 1975-1976  \\
\multicolumn{8}{c}{\emph{Observations in \cite{Bec1982}}} \\
 51 & \cite{Pickering1904a}              &     0/    1 &      ---   &	   ---   &      ---   &      ---   & 1904-1904* \\
 52 & \cite{Pickering1904b}              &     0/    1 &      ---   &	   ---   &      ---   &      ---   & 1904-1904* \\
 53 & \cite{Perrine1904}                 &     5/    5 &    -0.4413 &	 -0.1065 &     0.9992 &     0.5282 & 1904-1904* \\
 54 & \cite{Albrecht1909}                &    11/   11 &    -2.3742 &	  1.3441 &     2.5230 &     2.1368 & 1905-1905* \\
 55 & \cite{Perrine1909}                 &    10/   10 &    -1.0285 &	  0.0567 &     1.6793 &     0.7945 & 1906-1908* \\
 56 & \cite{MNRAS68-1908}                &    15/   15 &    -0.1002 &	  1.2292 &     0.6667 &     1.4391 & 1907-1907* \\
 57 & \cite{MNRAS69-1909}                &    23/   23 &     0.6177 &	  0.8746 &     0.8672 &     1.3982 & 1908-1908* \\
 58 & \cite{MNRAS70-1910}                &    12/   12 &     0.7487 &	  1.1878 &     1.3078 &     1.5321 & 1909-1910* \\
 59 & \cite{MNRAS71-1911}                &     7/    7 &     1.3857 &	  1.4758 &     1.8302 &     1.6171 & 1910-1910* \\
 60 & \cite{Barnard1913}                 &     2/    2 &     0.4634 &	 -0.4752 &     1.4150 &     0.5778 & 1912-1913* \\
 61 & \cite{Barnard1914}                 &     2/    2 &    -0.9066 &	 -0.5416 &     1.0271 &     0.5457 & 1913-1913* \\
 62 & \cite{VanBiesbroeck1922}           &     0/    5 &      ---   &	   ---   &      ---   &      ---   & 1922-1922* \\
 63 & \cite{Richmond1943}                &     1/    1 &    -0.5528 &	  0.0497 &     0.5528 &     0.0497 & 1940-1940  \\
 64 & \cite{VanBiesbroeck1944}           &     8/    8 &     1.4202 &	 -0.0666 &     1.8178 &     0.3589 & 1942-1942  \\
 65 & \cite{Bobone1953} 	          &     7/    7 &    -0.7151 &	 -0.0210 &     1.1774 &     0.3068 & 1952-1952  \\
 66 & \cite{VanBiesbroeck1957}           &    13/   14 &    -0.2220 &	  0.5337 &     0.9805 &     0.7550 & 1955-1955  \\
 67 & \cite{VanBiesbroeck1958}           &     8/    8 &    -0.6959 &	  1.1637 &     0.9347 &     1.3354 & 1957-1957  \\
 68 & \cite{Roemer1966}                  &     2/    2 &     1.0106 &	  0.3198 &     1.0171 &     0.3413 & 1960-1960  \\
 69 & \cite{VanBiesbroeck1976}           &     3/    4 &    -1.2365 &	  0.7450 &     2.6553 &     0.7780 & 1969-1969  \\
 71 & \cite{Debehogne1981a}              &    21/   21 &     0.5240 &	 -0.1129 &     0.6780 &     0.5557 & 1981-1981  \\
    & \cite{Debehogne1981b}             &      &      &	  &      &       &   \\
 72 & \cite{Bowell1981}                  &     8/    8 &     0.3358 &	 -1.5206 &     0.8243 &     1.6113 & 1981-1981  \\
 73 & Debehogne H. privat comm. 1981     &    18/   18 &    -0.2414 &	 -0.6121 &     0.5502 &     0.8142 & 1982-1982  \\
 74 & \cite{Dourneau1991}                &     5/    5 &     0.1846 &	 -0.2148 &     0.2326 &     0.3068 & 1989-1989  \\
\multicolumn{8}{c}{\emph{Observations in NSDC}} \\
102 & \cite{Ledovskaya1999}              &     9/    9 &     0.0327 &	 -0.3741 &     0.0795 &     0.3846 & 1998-1998* \\
103 & \cite{StoneHarris2000}             &    38/   38 &    -0.0901 &	 -0.0090 &     0.2336 &     0.3486 & 1998-1999  \\
104 & \cite{Stone2000}                   &    19/   19 &     0.0158 &	  0.0829 &     0.2770 &     0.2154 & 1999-1999  \\
105 & Derek Jones comm to NSDC           &     7/    7 &    -0.0385 &	  0.0701 &     0.1007 &     0.1803 & 1995-1997  \\
106 & \cite{Veiga2000}                   &    60/   60 &    -0.0745 &	  0.2488 &     0.1828 &     0.3986 & 1995-1997  \\
107 & MPC comm. (2001-2002a)             &    22/   22 &    -0.2601 &	  0.2854 &     0.4536 &     0.4563 & 2000-2000* \\
108 & MPC comm. (2001-2002b)             &     9/    9 &    -0.3174 &	  0.5602 &     0.4106 &     0.5755 & 2000-2000* \\
109 & MPC comm. (2001-2002c)             &     6/    6 &    -0.0137 &	 -0.0085 &     0.0318 &     0.0306 & 2001-2001* \\
110 & MPC comm. (2001-2002d)             &    18/   18 &     0.0451 &	  0.1962 &     0.4876 &     0.2463 & 2001-2001* \\
111 & MPC comm. (2001-2002e)             &     3/    3 &    -0.4311 &	 -0.1218 &     0.4765 &     0.3222 & 2001-2001  \\
113 & Comm. to NSDC (2002)               &     3/    3 &     0.5031 &	 -0.7376 &     0.7787 &     0.7420 & 2002-2002* \\
114 & MPC comm. (2002)                   &     3/    3 &    -0.0223 &	  0.1123 &     0.0453 &     0.1131 & 2002-2002* \\
115 & Comm. to NSDC (2003a)              &    22/   22 &     0.0162 &	 -0.4346 &     0.6929 &     1.2151 & 2002-2003* \\
116 & Comm. to NSDC (2003b)              &    38/   38 &     0.0143 &	  0.1562 &     1.1432 &     1.0772 & 1998-2003* \\
117 & Comm. to NSDC (2003c)              &    16/   16 &    -0.0630 &	  0.0440 &     0.0857 &     0.0564 & 2003-2003* \\
118 & \cite{Fienga2002}                  &   162/  162 &     0.0720 &	 -0.1076 &     0.1855 &     0.2209 & 1998-1999* \\
120 & Comm. to NSDC (2003d)              &     2/    2 &     0.0209 &	 -0.0003 &     0.0423 &     0.0036 & 2003-2003* \\
121 & Comm. to NSDC (2003-2004)          &    34/   34 &    -0.1025 &	  0.0402 &     0.3632 &     0.1568 & 2003-2003* \\
122 & Comm. to NSDC (2004)               &     5/    5 &    -0.0698 &	 -0.0109 &     0.0772 &     0.0693 & 2004-2004* \\
125 & \cite{Peng2006}                    &   210/  210 &     0.0269 &	 -0.0163 &     0.0592 &     0.0575 & 2003-2005* \\
127 & \cite{Qiao2006}                    &   115/  115 &    -0.0074 &	  0.0046 &     0.1191 &     0.1508 & 2003-2004* \\
137 & MPC 70653 WISE Observations        &     5/    5 &    -0.0593 &	  0.4536 &     0.1572 &     0.5128 & 2010-2010  \\
139 & \cite{Qiao2011}                    &  1173/ 1173 &    -0.0725 &	  0.0251 &     0.1486 &     0.1099 & 2005-2008* \\
200 & MPC24160                           &     4/    4 &    -0.6614 &	 -0.0325 &     1.3239 &     0.7528 & 1994-1994  \\
201 & MPC38967                           &     8/    8 &    -0.1696 &	 -0.0770 &     0.1981 &     0.0933 & 1999-1999  \\
202 & MPC40910                           &    38/   39 &     0.6800 &	 -0.1226 &     1.2349 &     1.1008 & 1995-1998  \\
203 & MPC40911                           &    17/   17 &    -0.1276 &	  0.0833 &     0.3540 &     0.3465 & 1998-2000  \\
204 & MPC41261                           &     3/    3 &    -0.1428 &	  0.3942 &     0.1443 &     0.3948 & 2000-2000  \\
205 & MPC41449                           &    10/   10 &    -0.1008 &	  0.1553 &     0.3738 &     0.3201 & 2000-2000  \\
206 & MPC41830                           &     3/    3 &    -0.2477 &	  0.3817 &     0.2626 &     0.4060 & 2000-2000  \\
207 & MPC43757                           &     8/    8 &     0.3832 &	  0.8383 &     1.3833 &     1.3944 & 2001-2001* \\
208 & MPC44182                           &    11/   11 &     0.4054 &	 -0.1473 &     0.6700 &     0.3502 & 2001-2001* \\
209 & MPC46761                           &     6/    6 &     0.0458 &	  0.0652 &     0.0739 &     0.0988 & 2002-2002* \\
210 & MPC50594                           &     7/    7 &     0.0202 &	  0.0621 &     0.0927 &     0.1233 & 2004-2004* \\
211 & MPC51181                           &     3/    3 &     0.0779 &	  0.0451 &     0.0801 &     0.0453 & 2004-2004  \\
212 & MPC51367                           &     6/    6 &     0.0582 &	  0.0325 &     0.0659 &     0.0418 & 2004-2004  \\
213 & MPC51497                           &     4/    4 &     0.0246 &	  0.0406 &     0.0403 &     0.0412 & 2004-2004* \\
215 & MPC53171                           &     3/    3 &    -0.1010 &	 -0.0407 &     0.1257 &     0.0479 & 1996-1996* \\
216 & MPC53464                           &    19/   19 &    -0.0555 &	 -0.0195 &     0.1914 &     0.1828 & 2004-2005  \\
217 & MPC53629                           &     8/    8 &    -0.0015 &	 -0.0514 &     0.1990 &     0.1518 & 2005-2005  \\
218 & MPC53948                           &    36/   36 &     0.0253 &	 -0.0210 &     0.2774 &     0.2015 & 2004-2005* \\
219 & MPC54163                           &    22/   22 &     0.0228 &	  0.0053 &     0.2475 &     0.1784 & 2004-2005* \\
220 & MPC55508                           &     5/    5 &    -0.0147 &	  0.8286 &     0.2316 &     0.8815 & 2005-2005* \\
221 & MPC55975                           &     6/    6 &     0.3247 &	 -0.0217 &     0.7387 &     0.2284 & 2005-2005* \\
222 & MPC55976                           &    72/   72 &     0.0081 &	  0.0544 &     0.2003 &     0.2815 & 2005-2006* \\
223 & MPC56148                           &    34/   34 &     0.0053 &	  0.0140 &     0.2814 &     0.2085 & 2006-2006* \\
224 & MPC56608                           &    33/   33 &     0.1470 &	 -0.0966 &     0.2252 &     0.1823 & 2006-2006* \\
225 & MPC56609                           &     1/    1 &     0.2920 &	 -0.2526 &     0.2920 &     0.2526 & 2006-2006* \\
226 & MPC57946                           &     7/    7 &    -0.5558 &	  0.6623 &     0.8614 &     1.5035 & 2006-2006* \\
227 & MPC57947                           &     3/    3 &     1.5135 &	  1.2146 &     1.9990 &     1.2276 & 2006-2006* \\
228 & MPC58097                           &    10/   10 &     0.0283 &	 -0.1981 &     0.2801 &     0.4356 & 2006-2006* \\
229 & MPC58523                           &    12/   12 &     0.0619 &	 -0.0699 &     0.5181 &     0.2336 & 2006-2006* \\
230 & MPC58764                           &    34/   34 &     0.2153 &	  0.0840 &     0.4006 &     0.6965 & 2007-2007* \\
231 & MPC59029                           &     9/    9 &    -0.0607 &	  0.1404 &     0.2501 &     0.5894 & 2007-2007* \\
232 & MPC59305                           &    23/   23 &     0.1357 &	  0.0140 &     0.2595 &     0.2105 & 2007-2007* \\
233 & MPC59581                           &    18/   18 &     0.1645 &	 -0.0935 &     0.2798 &     0.1918 & 2007-2007* \\
234 & MPC59860                           &    17/   17 &     0.2167 &	 -0.1131 &     0.3165 &     0.2162 & 2007-2007* \\
235 & MPC60086                           &    14/   14 &     0.0379 &	  0.0344 &     0.3345 &     0.2506 & 2007-2007* \\
236 & MPC61685                           &     1/    1 &     1.2399 &	 -0.2102 &     1.2399 &     0.2102 & 2008-2008* \\
237 & MPC61686                           &     3/    3 &     0.3235 &	 -0.1693 &     0.4012 &     0.2109 & 2008-2008* \\
238 & MPC61978                           &    30/   30 &    -0.0142 &	  0.0457 &     0.2321 &     0.2438 & 2008-2008* \\
239 & MPC62254                           &    20/   20 &     0.0054 &	  0.1463 &     0.3018 &     0.4611 & 2008-2008* \\
240 & MPC62566                           &    23/   23 &     0.0406 &	 -0.0538 &     0.2190 &     0.2306 & 2008-2008* \\
241 & MPC62864                           &     8/    8 &     0.2757 &	  0.1044 &     0.4027 &     0.1738 & 2008-2008  \\
242 & MPC62865                           &     2/    2 &     0.4133 &	 -0.0853 &     0.4201 &     0.0967 & 2008-2008  \\
243 & MPC63123                           &    14/   14 &     0.3811 &	 -0.2029 &     0.4459 &     0.2786 & 2008-2008  \\
246 & MPC63584                           &     3/    3 &     0.2071 &	  0.2070 &     0.3299 &     0.2858 & 2008-2008  \\
247 & MPC64482                           &     4/    4 &    -0.0818 &	  0.3491 &     0.4814 &     0.5066 & 2008-2008* \\
248 & MPC64749                           &     6/    6 &     0.0161 &	  0.5311 &     0.2223 &     0.6249 & 2008-2008* \\
249 & MPC65036                           &     9/    9 &     0.0818 &	  0.0208 &     0.1599 &     0.0503 & 2009-2009* \\
250 & MPC65325                           &     4/    4 &    -0.5536 &	  0.2077 &     0.6222 &     0.2978 & 2009-2009  \\
251 & MPC65628                           &    15/   15 &     0.1077 &	 -0.0344 &     0.2934 &     0.2110 & 2009-2009  \\
252 & MPC65920                           &    16/   16 &    -0.1531 &	  0.3044 &     0.5297 &     0.8047 & 2009-2009* \\
253 & MPC66188                           &     2/    2 &    -0.5985 &	  0.0678 &     0.6111 &     0.3130 & 2009-2009  \\
254 & MPC66450                           &     6/    6 &     0.0406 &	  0.0375 &     0.2085 &     0.1453 & 2009-2009* \\
255 & MPC66686                           &     4/    4 &     0.1597 &	 -0.1154 &     0.2672 &     0.2136 & 2009-2009  \\
257 & MPC68212                           &     2/    2 &    -0.0147 &	 -0.1053 &     0.0530 &     0.1070 & 2010-2010* \\
258 & MPC68669                           &     6/    6 &    -0.0511 &	  0.0803 &     0.2327 &     0.1850 & 2010-2010* \\
259 & MPC69202                           &    30/   30 &     0.0601 &	  0.3838 &     0.3508 &     0.8498 & 2010-2010* \\
260 & MPC69724                           &    21/   21 &     0.6952 &	  0.9287 &     0.9956 &     1.3787 & 2010-2010* \\
261 & MPC70193                           &     4/    4 &     0.0733 &	  0.1649 &     0.1064 &     0.1968 & 2010-2010* \\
262 & MPC71062                           &     1/    1 &     0.0851 &	  0.1336 &     0.0851 &     0.1336 & 2010-2010* \\
263 & MPC73667                           &     2/    2 &    -0.0898 &	  0.0278 &     0.1128 &     0.0305 & 2011-2011* \\
264 & MPC74025                           &    21/   21 &     0.0002 &	 -0.0429 &     0.1241 &     0.1678 & 2011-2011* \\
267 & MPC74380                           &    39/   39 &     0.2880 &	  0.1457 &     0.5760 &     0.5211 & 2011-2011* \\
268 & MPC74811                           &    29/   29 &     0.1555 &	  0.0254 &     0.3181 &     0.1771 & 2011-2011* \\
270 & MPC75400                           &     3/    3 &     0.2502 &	  0.0030 &     0.2608 &     0.0727 & 2011-2011* \\
271 & MPC75593                           &     9/    9 &     0.3081 &	 -0.0401 &     0.3166 &     0.1026 & 2011-2011* \\
272 & MPC75849                           &     6/    6 &     0.0941 &	 -0.2012 &     0.1434 &     0.2334 & 2011-2011* \\
\multicolumn{8}{c}{\emph{Other observations}} \\
300 & \cite{Pickering1908}               &     6/   42 &     0.2576 &	  0.9413 &     1.9223 &     3.0575 & 1898-1902* \\
301 & \cite{NOFS}                        &   205/  205 &     0.0201 &	  0.0420 &     0.2786 &     0.3851 & 2000-2011* \\
350 & Voyager data - \cite{Jacobson1998} &     8/    8 &    -6.6018 &	 -1.1079 &     6.6194 &     1.2836 & 1981-1981  \\
    & \cite{Jacobson1998}  & & & & & &  \\
351 & Cassini data                       &   223/  223 &     0.6304 &	 -0.1291 &     2.5353 &     1.5782 & 2004-2004  \\
    & Tajeddine R., privat comm. 2012  & & & & & &  \\
500 & Minor Planet Center (2011)         &    83/   83 &     0.1518 &	  0.1092 &     0.4838 &     0.3341 & 1907-2012* \\
 \hline
\hline
\end{longtable}
\end{small}

\end{document}